# Fermi surface map of the single-layer Bi-cuprate $Bi_2Sr_{2-x}La_xCuO_{6+\delta}$ at optimal doping

[a] R. Müller, [a] M. Schneider, [a] C. Janowitz, [a] R.-St. Unger, [a] T. Stemmler, [a] A. Krapf, [a] H. Dwelk, [a] R. Manzke, [b] K. Roßnagel, [b] L. Kipp, [b] M. Skibowski

A first Fermi surface map of a single-layer high-$T_c$ superconductor is presented. The experiments were carried out on optimally doped $Bi_2Sr_{2-x}La_xCuO_{6+\delta}$ (x=0.40) with synchrotron radiation which allow to discuss in detail the strong polarisation dependence of the emissions near the Fermi edge. For the cuprates only little is known about the impact of the electron-photon matrix element determining the photoelectron intensity. For the example of the model layered superconductor $Bi_2Sr_{2-x}La_xCuO_{6+\delta}$ it will be demonstrated that the polarization geometry has significant influence on the energy distribution curves at $E_F$ and consequently also for the determination of the topology and character of the Fermi surface (FS) by angle-resolved photoemission. For further clarification also a FS map of the n=2 material Bi-2212 has been measured applying a different polarisation geometry as previously used by Saini et al.. In the context of the current debate on the character of the Fermi surface of Bi-cuprates our results confirm a hole-like Fermi surface for n=1 as well as n=2 material, what might be the universal FS for high-$T_c$ superconductors.

[a] Institut für Physik, Humboldt-Universität zu Berlin, Invalidenstraße 110 , D-10115 Berlin, Germany
[b] Institut für Experimentelle und Angewandte Physik, Universität Kiel, D-24098 Kiel, Germany

## I.     INTRODUCTION

Angle-resolved photoemission spectroscopy (ARPES) is a powerful tool to study the electronic structure of high temperature superconductors (HTSC) and other strongly correlated systems since it allows to probe the energy and momentum relations directly. The progress achieved by ARPES over the last decade in order to understand the characteristics in the electronic structure of high-$T_c$ superconducting cuprates is impressive and important results have been obtained from both the normal and the superconducting state. As examples may be mentioned the investigations of the weakly dispersing Cu-O derived Zhang-Rice (ZR) singlet band and the hole-like Fermi surface (FS), the observation of pseudo gaps, and the anisotropy of the superconducting gap. [1]

The Fermi surface is one of the most important normal state properties. Although the studies started already in the early 90th, the Fermi surface is still an object of heavy debate. Here are two questions important: What is the character (or topology) and is the topology universal for high-$T_c$ superconductors? Regarding the character of the Fermi surface, the accepted picture after the detailed work of Ding et al. [2] on the double-layer (n=2) material of BSCCO was that it is hole-like with a topology of pockets centred at the Brillouin zone corners (X,Y-points [3]). In the following, the hole-like Fermi surface has been often confirmed but has been challenged recently by a study of Chuang et al. [4]. These authors mentioned the strong photon energy dependence of the spectral features at the Fermi level and interpreted the measurements at hν=32-33 eV as due to an electron-like Fermi surface with a large electron pocket centred at Γ. After this, many investigations were initiated in order to find an explanation for the contradictory results. Borisenko et al. [5] presented a remarkably complete Fermi surface map of the double-layer material Bi-2212 applying HeI radiation (hν=21.22eV), in tradition of the famous data of Aebi et al. [6] who introduced this new ARPES method. Besides the FS due to the main band (MB) they were able to resolve also the details of the surfaces resulting from the umklapp bands (or diffraction replica DR) and the shadow bands (SFS). In particular, Borisenko et al. [5] emphasized the importance of polarisation effects and the resulting intensity variations of the emissions at $E_F$. As a consequence, they meant, the presented details of the FS became only possible with non-polarized radiation, an advantage for measurements with the He resonance lamp. Another measurements of Bi-2212 taken with polarized synchrotron radiation [7, 8] by the traditional photoemission technique (energy distribution curves (EDC) along different high symmetry lines of the Brillouin zone and looking for crossings of the Fermi energy) confirmed the strong polarization dependencies. Taking polarization effects and the resulting influences on the transition matrix elements more properly into account these investigations [7, 8] showed again that the Fermi surface of Bi-2212 is hole-like. Nevertheless, these results are again questioned by others [9-12], leaving the important question on the Fermi surface up to now unsolved when judging the accumulated results of Bi-2212. First investigations on the FS of a single-layer material, Bi-2201, with the traditional photoemission technique has been published by Müller et al. [13] , Mesot et al. [14] and Sato et al. [15]. The authors found that the crossings with $E_F$ (or Fermi wave vectors $k_F$) are in close correspondence to them observed in Bi-2212 and in agreement with a hole-like Fermi surface.

In the present contribution we report, for the first time, a Fermi surface map of single-layer material, namely Bi-2201. The single crystals were optimally doped ($T_c$=29K) according to the formula $Bi_2Sr_{2-x}La_xCuO_{6+\delta}$ with x=0.40. Our aim here is to show, besides the presentation of an almost complete Fermi surface of a single-layer material, also the unique advantages of Fermi surface mapping with synchrotron radiation. Thus, the key problems of this method when applied on cuprates will be discussed which arise from the strong photon energy dependencies and polarization effects of the emissions at $E_F$. These lead to severe suppression or enhancement of the intensity at certain values of the huge parameter space accessible with photoemission [1, 16, 17].

Even more, by mapping additionally the double-layer material Bi-2212 at clear polarisation geometries, which have been not applied previously, we will show that all discrepancies observed between maps taken with synchrotron and HeI radiation can be consistently explained by taking the polarisation of the light properly into account.

## 2. EXPERIMENTAL

Our group is able to grow single crystals of high quality of the first three members of the Bi-based cuprate family $Bi_2Sr_2Ca_{n-1}Cu_nO_{2n+4+\delta}$ with n = 1, 2, and 3. In addition, the hole concentration of the samples can be varied almost continuously by substituting e.g. trivalent Yttrium or Lanthanum [18]. For the present experiments optimally doped n=1 and slightly overdoped n=2 material is chosen. The advantage of the single-layer n=1-material in comparison to n=2 or 3 is the simple fact that for the superconducting as well as the normal state the properties are not affected by possible bilayer splitting effects. Furthermore, the low $T_c$ of the single-layer material Bi-2201 makes it possible to study the electronic structure of the normal state at about 30K, i.e. with a much better energy resolution than for n=2 Bi-2212 due to much lower thermal broadening.

The optimally doped $Bi_2Sr_{2-x}La_xCuO_{6+\delta}$ single crystals with x=0.40 and $T_c$=29K were grown out of the stoichiometric melt where we varied the conditions given by Nameki et al. [19] only slightly. Lanthanum-free samples are strongly hole-overdoped with a $T_c$ of about 7 K. In order to reach optimal $T_c$ part of the $Sr^{2+}$ is replaced by $La^{3+}$ what reduces the hole concentration in the Cu-O plane. The dependence of $T_c$ on x is about parabolic. Optimal $T_c$ of about 29 K and sharpest transition $\Delta T_c$ of 2 K is found for a La content of x=0.4. The transport properties and the chemical composition of the samples were controlled by measuring the ac susceptibility and x-ray emission spectra (EDX), respectively (the details will be published elsewhere [18]). The $Bi_2Sr_2CaCu_2O_{8+\delta}$ single crystals were grown conventionally out of the solution (Bi-rich melt). The pristine samples were slightly overdoped with a $T_c$ of 89K. All samples were rectangular shaped with the long side along the crystallographic a-axis, as confirmed by Laue diffraction and *in situ* electron diffraction (LEED). Typical sizes of the samples are 5mm×2mm.

The angle-resolved photoemission data were taken at the synchrotron radiation laboratory HASYLAB in Hamburg with the hemispherical deflection analyser ASPHERE of the University of Kiel mounted on a two-axes goniometer. We used 18 eV photon energy from the high-resolution 3m normal-incidence monochromator HONORMI. The overall energy resolution was 40 meV, determined from the Fermi edge of evaporated gold at low temperatures. The acceptance angle was 1°. In order to perform Fermi surface maps the ASPHERE spectrometer has been improved to enable motorized and computer controlled sequential angle-scanning for data acquisition [20]. Therefore, with this system Fermi surface mapping becomes possible without moving the sample. The advantage is twice: At first, cooling of the samples on a stationary He evaporation cryostat turns out to be quite easy and, at second, particular and clear polarisation geometries can be much better controlled than in a system where the samples must be moved with respect to the incident synchrotron beam. For the Fermi surface maps shown here the energy of the electron analyser is fixed at the Fermi level $E_F$ and the photoemission intensity is detected as a function of the emission angles. The energy window was again 40 meV. At k-locations on the Fermi surface high photoemission intensity is expected. The intensities of all spectra and Fermi surface plots were normalized at a binding energy of 500 meV as common. At this energy the photoemission signal is featureless in energy and momentum.

Additional polarisation dependent photoemission spectra were taken with 34 eV photon energy at the combined undulator-monochromator beamline U2-FSGM [21] at BESSY, Berlin. The crossed undulator U2 allowed a rotation of the electrical field vector **E** by 90°. These are optimal conditions because it allows polarisation dependent measurements without moving the sample. It has been tested by determining the Fermi-Dirac distribution of a gold film in electrical contact with the sample that uncertainties due to photon energy changes between both polarizations are below 2 meV. This serves also for the determination of the Fermi energy $E_F$. At BESSY we used the high-resolution angle-resolving photoemission station HIRE-PES, also equipped with a spherical analyser on a two-axes goniometer [22]. The resolution of this system has been tested to reach 8 meV demonstrated with synchrotron radiation at the Fermi edge of gold at low temperature. For the spectra shown here the overall energy resolution was 30 meV and the electron acceptance angle was 1°.

## 3. FERMI SURFACE MAPS
### 3.1 Results of n=1 $Bi_2Sr_{2-x}La_xCuO_{6+\delta}$, x=0.40

In Figure 1a the $\mathbf{k}_{\|}$-map of the intensity of the emitted photoelectrons from optimally doped single-layer $Bi_2Sr_{2-x}La_xCuO_{6+\delta}$ single crystals is shown for the normal state at 40K. The electrons were collected within a 40 meV wide energy window centered at the Fermi energy. Into the experimental map the theoretical Fermi surface given for the double-layer material Bi-2212 by Ding et al. [2] is drawn for comparison. Directly observable are at this temperature all expected parts of the Fermi surface, due to the main band (MB, thick lines), the umklapp bands or diffraction replica of the main band (DR, dotted lines), and the shadow bands (SFS, dashed lines). The SFS is visible as a weaker feature near the DR surface on the direction to the Y-point.

The traditional way to determine Fermi surfaces was to measure energy distribution curves (EDC) for different k-points of the Brilliouin zone and to determine the k-locations where bands cross the Fermi energy. Naturally, due to the large number of spectra needed by this method only a few k-locations on the Fermi surface are usually probed. On the other hand, this method allows a very precise determination of the Fermi wave vector $\mathbf{k_F}$ [23]. Figure 1b shows such an EDC series along the crystallographic direction ΓY taken on a similar optimally doped sample. From the spectra, one can easily obtain the first DR-crossing at 21% of the distance ΓY and the crossing of the MB at 40%ΓY. The situation near 65% ΓY where the second DR band and the SFS band are expected (see Figure 1a) is more complex. The FS features DR and SFS overlap here and the EDC series reveals only a broad crossing region. More insight in this detail of the k-space can be gained from a measurement of the momentum distribution curve (MDC). MDCs are detected at a fixed kinetic energy, e.g. at the Fermi energy, while scanning the emission angle along certain direction of the Brillouin zone. A MDC of n=1 Bi-2201 is shown along the ΓY direction in the right panel of Figure 1b. Here the advantages of MDCs are obvious. Besides that one gets the FS information in much faster time, FS crossings can be detected with distinctly increased resolution. In the MDC of Figure 1b the crossings of the SFS and DR band are resolved what, on the other hand, also demonstrates the good resolution of the FS map of Figure 1a. The DR bands and the corresponding DR Fermi surface are caused by umklapps of the main band due to the approximate 5x1-superstructure along the ΓY direction. The diffraction vector is m$\mathbf{q}$=(0.21π, 0.21π), whereby m is the order of DR. The origin of the SFS Fermi surface is still in discussion but most probably it seems to be driven by antiferromagnetic correlations [2, 5, 6].

In line with symmetry selection rules, discussed in more detail below, the emissions from the main band are suppressed in polarisation geometries where the planes of electron detection $\mathbf{p}$ and vector potential $\mathbf{E}$ of the synchrotron light become coplanar. This situation is encountered in the lower-left quadrant (or X-quadrant) of the FS map of Figure 1a. The intense emission for all observed bands along the ΓY direction is therefore ascribable to the used polarisation geometry prevailing in the hole lower-right quadrant. Here $\mathbf{p}$ and $\mathbf{E}$ are almost perpendicular to each other.

Three experimental observations should be mentioned briefly with respect to previous results (though on the double-layer material Bi-2212): (i) The intensities of parts of the Fermi surface of the Bi compounds vary very strongly has already been observed by Saini et al. [24, 25] on n=2 Bi-2212. In their experiments they applied a polarisation geometry different to ours, with the result that the part of low intensity was in their geometry the Y quadrant. They claimed that the observation of missing segments of the Fermi surface along the ΓY direction may be a hint towards a striped phase. (ii) In our polarisation geometry the SFS is observable in the ΓY direction that does also show up in studies of Bi-2212 with non-polarized HeI radiation [5, 6]. (iii) In contradiction to the Fermi surface map of Saini et al. [24, 25] for Bi-2212 and our result shown in Figure 2, for Bi-2201 almost no spectral weight of the main band is observed near the M-point (Figure 1a). The weak intensity between Γ and M is most probably due to the crossing of DR bands in this k-region predicted also by Ding et al. [2].

A detailed explanation of the experimental observations by variations of the transition matrix elements will be given in chapter 4.

## 3.2   Results of n=2 $Bi_2Sr_2CaCu_2O_{8+\delta}$

As mentioned in the introduction, most of our knowledge on Fermi surfaces of Bi-based cuprates comes from measurements of the double-layer system $Bi_2Sr_2CaCu_2O_{8+\delta}$ [2, 5, 6]. The most complete Fermi surface maps were taken with non-polarized HeI radiation, establishing a hole-like Fermi surface [5, 6]. There is only one map for the Bi-compound were synchrotron radiation has been used [24, 25]. With the linearly polarised synchrotron radiation the authors observed characteristic "missing segments" of the Fermi surface and other locallized areas of suppressed intensity - the so-called "hot spots". In order to test our experimental set-up and to repeat the previous measurements on n=2 material with a partly different polarisation geometry we performed a Fermi surface map on a slightly overdoped $Bi_2Sr_2CaCu_2O_{8+\delta}$ single crystal with $T_c$=85 K. With this measurement we follow an idea and theoretical prediction of Mesot et al. [26]. The samples were at room temperature during the measurements.

In Figure 2 the measured Fermi surface map of Bi-2212 is shown. The theoretically expected Fermi surface [2] is again added to the experiment. The high intensities (white regions) result at $k_\parallel$ locations where bands cross the Fermi energy. The sample was aligned in ΓM direction giving a polarisation geometry where the Cu-O bonds are parallel and perpendicular to the electrical field vector **E** of the synchrotron light (see also Figure 3). In contradiction to Saini et al. [24, 25] we did not rotate the sample. In our case the crystal is fixed and the motorized spectrometer scans the Fermi surface over two independent goniometer axes. Thus, our polarisation geometry is not the same as in the work of Saini et al. [24, 25] but, in principal, similar and with respect to particular high-symmetric directions identical.

In the experimental map of Figure 2 the Fermi surface of the Cu-O derived main band (MB) is observable in both quadrants, although in the Y-quadrant much weaker than in the X-quadrant. The additional features like the diffraction replica (DR) of the main FS and the shadow Fermi surface (SFS) are not visible in the map under the applied polarisation geometry. Note, in addition, the close correspondence between the experimental Fermi surface map and the one calculated for identical polarisation conditions using relations given by Mesot et al. [26]. It is shown in the right panel of Figure 2. This is also in close agreement with the results around the M-point of Saini et al. [24, 25] as expected for the horizontal high symmetry direction ΓM. Regarding the relative intensities of the Fermi surface of the main band (MB), it is larger in the X-quadrant than in the Y-quadrant, whereas the highest intensity appears near the M point. This could be either caused by the DR band [26] or new additional band along this ΓM direction [25]. Although we agree with Saini et al. [24, 25] that the intensity is very high near M, in particular in the X-quadrant, it is important to note that we have no hints for a missing segment along ΓY. The weak intensities in the Y-quadrant and, in particular, along the ΓY direction are explainable, similarly to Bi-2201, by variations of the transition matrix element, see chapter 4.

## 4. DISCUSSION

In order to get more insight into the peculiar polarisation effects of the Bi-superconductors we performed additional high-resolution ARPES spectra of optimally doped $Bi_2Sr_{2-x}La_xCuO_{6+\delta}$ single crystals at specific points of the Brillouin zone. A previous study of n=1 material performed with synchrotron radiation [28] has not accounted for the differences between the ΓY and ΓX direction. Because of the superstructure arising from the slight mismatch between BiO and $CuO_2$ planes due to excess oxygen, the ΓY and ΓX directions are not equivalent. As a consequence, the umklapp bands due to the superstructure are clearly observed in Figure 1 in the ΓY direction, in line with experimental results of n=2 material [2, 7]. These umklapp emissions show up in photoemission either by the photoelectrons being diffracted by the superlattice (SL) or by the BiO superlattice distortion affecting directly the electronic structure of the $CuO_2$ planes. Therefore, the polarisation dependent spectra shown in Figure 4 were taken near the Fermi surface crossing of the main band of about 0.4ΓX and 0.4ΓY, not to be influenced by the superstructure. In one polarisation geometry **E** is parallel to the detection plane of the photoelectrons, in the other perpendicular to it. This was achieved here by rotating the sample by 90°. It results that for both directions, ΓY and ΓX, one measures less intensity at $E_F$ if the electron emission **p** is in the mirror plane defined by the surface normal and the polarization direction (Figure 4). This intensity variation is in agreement with observations of Bi-2212 [1]. It can be described by the variation of the transition matrix element at these controlled polarisation conditions predicting an initial state of d $x^2-y^2$ symmetry on a Cu-site (see Figure 3).

While the photoemission intensities can be studied precisely within the one step model and a complete band structure calculation such as LDA band theory [29], we analyse in the following the effects of the transition matrix element only qualitatively. Here we follow the ansatz given by Gobeli et al. [30] and Dietz et al. [31] which has been also successfully applied to the high-$T_c$ superconductors [1]. ARPES probes the occupied part of the electronic states. The intensity I(**k**,ω) is proportional to the square of the transition matrix element $|M_{if}(\mathbf{k})|^2$, the Fermi function f(ω), and the spectral function A(**k**,ω) of the initial state. The dipole part of the transition matrix element is given by $M_{if}(\mathbf{k})=(e/mc)\langle\Psi_f(\mathbf{k})|\mathbf{A}\cdot\mathbf{p}|\Psi_i(\mathbf{k})\rangle$, where the vector potential of the light **A** is colinear with the electrical field vector **E**. $\Psi_i(\mathbf{k})$ and $\Psi_f(\mathbf{k})$ are the wave functions of the initial and final states, respectively. The essential point is that, since $M_{if}(\mathbf{k})$ is the observable quantity, it must be invariant under symmetry operations, i.e. for the polarisation geometry applied in the photoemission experiment. Only those combinations of the initial state, the final state, and the vector potential contribute to the photoemission process which leave $M_{if}(\mathbf{k})$ invariant.

In the case of the photoemission spectra of Figure 4 the polarization vector of the incoming synchrotron radiation was aligned with the **a** or **b** axis of the sample corresponding to the ΓX and ΓY direction, respectively. This will put the lobes of the Cu 3d $x^2-y^2$ orbital due to $\Psi_i(\mathbf{k})$ at 45° to the horizontal and vertical direction, as illustrated in the upper part of Figure 3. With respect to the ΓX and ΓY mirror planes, the initial state has odd symmetry and the final states approximated by outgoing plane waves have even symmetry relative to the mirror plane. Thus, for the spectra of Figure 4 the matrix element $M_{if}(\mathbf{k})$ is ultimately determined by the symmetry of the momentum operator **A**·**p**. For a non-vanishing matrix element, according to $M_{if}(\mathbf{k})$=even*odd*odd=even, an odd dipole operator is required what is fulfilled by choosing the emission direction **p** of the outgoing electrons in a plane perpendicular to the vector potential of the incoming photon field **A** (or the polarisation vector **E**). Emission within the ΓX(Y) mirror planes is not allowed because of even symmetry for the dipole operator and $M_{if}(\mathbf{k})$=even*odd*even=odd. Although one does not expect such strict situations as discussed above, this is in fact clearly observed in the spectra of

Figure 4. One may note also that the polarization effects are much more pronounced for spectra of the ΓY direction than for ΓX. This might be due to the fact that near the Fermi level crossing along ΓX the main band is superimposed by the DR bands (see Figure 2) for which along this direction the selection rules are not valid. The distinctly broader spectral line at $0.4\Gamma X_2$ than at $0.4\Gamma Y_2$ additionally supports this explanation. Our analysis of the polarisation dependence along the ΓX(Y) directions of optimally doped Bi-2201 is in agreement with the observations of n=2 material, Bi-2212, by Shen and Dessau [1]. The dependencies up to here are therefore explainable by dipole selection rules.

The polarization geometry of the spectra in Figure 4 is principally the same as for the Fermi surface map of Figure 1. The electrical field vector **E** lies within the mirror plane ΓX but for the FS map the detection plane is varied between perpendicular and parallel. Taking the dipole selection rules discussed above into account this explains directly the high intensity of the Fermi surfaces of the main and diffraction replica band in the right Y-quadrant of the map. The condition is similar to that of the $0.4\Gamma Y_2$-spectrum of Figure 4. On the other hand, in the left X-quadrant one expects weak intensities. This explains possibly already the decrease in intensity around the M-point of Figure 1 (see also the detailed discussion below). The strong polarisation effects for the ΓY directions may also provide an explanation for the so-called missing segments in the Fermi surface of high-$T_c$ superconductors stressed by several authors [32].

Coming back to the region around the M-point of the FS map of Figure 1. With the polarisation set-up of Figure 3, upper part, and measuring along the high symmetry line ΓM one has a mixed polarisation case with respect to $\Psi_i(\mathbf{k})$. In Figure 5, spectra taken at the M-point are shown for different polarisation geometries. The spectrum **E**∥ΓX of the lower panel of Figure 5 corresponds to the situation of Figure 1 and emphasizes further the weak intensity at this M-point. In particular important is the dramatic polarisation effect at M when the electrical field vector **E** is turned by 90°. Here the direction of **E** has been changed directly by the undulator, i.e. without moving the sample or the analyzer. It is important to note that the observed intensity variation at the M-point violates the discussed dipole selection rules of $M_{if}(\mathbf{k})$, which is identical in both geometries. A critical analysis of the spectral contribution at $E_F$ by Manzke et al. [15] reveals in addition an energy shift between the two polarisations, what means that the emission contribution consists of two spectral lines. By normalizing the spectra absolutely by the photon flux, one finds almost vanishing polarisation dependence for the line at higher binding energy (evident also from the coinciding emission background around 400 meV binding energy). The polarisation acts obviously very strongly on the spectral line at $E_F$, almost switching it on and off. It is clear that these experimental results at $E_F$ need a new explanation of the CuO-derived Zhang-Rice singlet band of the superconducting cuprates along ΓM. Very recently, the dispersion of the two spectral lines has been also investigated [33]. Regarding these details, the strong intensity decrease caused by the dramatic polarisation dependencies of the spectral line at $E_F$ should be responsible for the missing parts of the main band Fermi surface in the X-quadrant and the reduction around the M-point of Figure 1.

In order to derive again at clean polarisation geometry also for the M-point one has to rotate the sample by 45°. By this the lobes of the Cu 3d $x^2-y^2$ orbitals, the initial state, are brought parallel to the horizontal or vertical ΓM mirror planes. This sample alignment has been often used in ARPES studies of cuprates [7,8] and we applied it also for the Fermi surface map of n=2 Bi-2212 shown in Figure 2. With respect to both planes $\Psi_i(\mathbf{k})$ has now even symmetry. As the final states are still even, the matrix element $M_{if}(\mathbf{k})$ becomes even for a detector position near M(π,0) and odd near M(0,π). In contrast to the ΓX(Y) alignments discussed above, suppression and enhancement is now expected for a detection plane perpendicular and parallel to the ΓM mirror plane, respectively. This is in fact observed in the spectra shown in the upper part of Figure 5. The intensity variation is very strong and, therefore, measurements with samples aligned in ΓM direction should be treated very carefully. But again, these variations of the intensities at M are explainable by dipole selection rules.

Regarding the experimental Fermi surface map of Bi-2212 shown in Figure 2 a similar polarisation dependence applies to as for the n=1 Bi-2201 crystal. By means of the spectra of Figure 5, upper panel, the intensity from the main band varies distinctly in the different parts of the map: Near the left M-point (horizontal detection plane) the intensity is large and in the direction to the M-point at the bottom (vertical detection plane) it is weak, both in line with dipole selection rules. In the upper left quadrant, the Fermi surface of the main band is principally observable, but its intensity is very weak due to the same reasons. Because this is now the Y-quadrant where the emissions of the main band and the diffracted bands should be dipole forbidden, the parts from the diffraction replica are absolutly not seen. In summary, the variation of the transition matrix element explains easily the experimental observations: large and vanishing intensities around the horizontal and vertical M-points, respectively, vanishing intensities along the ΓY directions, and weak but not vanishing intensities along the ΓX directions. All this is also observable in the FS map of Saini et al. [24, 25, 27], although they rotated the sample for the measurement. The parts of the Fermi surface of weak intensity along the ΓY directions are often called missing segments. It has been demonstrated by our analysis of Bi-2212 and Bi-2201 that these intensity variations are pricipally explainable by dipole selection rules.

Beyond the polarisation effects due to changes of the dipole matrix element one observes in the experimental FS map around the left M-point a remarkable intensity variation: the intensity is increased in the X-quadrant and decreased in the Y-quadrant. Moreover in the Y-quadrant of the M-point the intensity is even more suppressed than expected by the calculation shown in the right panel of Figure 2 which account for the matrix element effects. This local suppression of spectral weight has been for the first time observed by Saini et al. [24,25]. It is located at k-points where the Fermi surfaces of the main band and the shadow band cross. These points of missing spectral weight on the Fermi surface are the so-called hot spots and are interpreted as strong hints in the electronic structure for stripe formation along ΓX (for details see [27]). Similar this missing intensity at one side of the M-point is also be observed for Bi-2201 (Figure 1) and is actually not explainable by selection rules. Interestingly, the missing intensities due to the hot spots are obviously not observable in Fermi surface maps taken with non-polarized HeI radiation [5]. Thus their origin might again be a polarisation effect. On the other hand, most of work by Borisenko et al. [5] is performed on Pb-substituted Bi-2212 in order to reduce the diffraction replica, what might have influence on the result.

## 5. SUMMARY

In this paper a Fermi surface map of the single-layer superconductor Bi-2201 and, for a new polarization geometry, that of the double-layer Bi-2212 are presented. Both maps reveal all expected features of the Fermi surface of Bi-cuprates, which are due to the main band, the umklapp bands or diffraction replica of the main band, and the shadow bands. The intensities are strongly polarisation dependent. The main polarisation dependencies can be consistently explained by variations of the transition matrix element. But beyond these strong dependencies there are additional effects present, like the hot spots near M or the strange polarization dependence of the new double-structure of the CuO-derived Zhang-Rice singlet band near M, which are not explainable by dipole selection rules. Obviously all the obscure polarisation effects are due to states located near the M-point of the Fermi surface.

The main polarisation dependencies indicate that the initial states have odd symmetry with respect to the ΓY and ΓX planes, respectively even symmetry with respect to the ΓM plane, and the orbital character of the CuO-derived band at $E_F$ exhibits significant d $_{x^2-y^2}$ symmetry. This is in good agreement with x-ray absorption measurements (see e.g. [17]) and theoretical models of the electronic structure of the cuprates [34]. It should be noted that strict selection rules could not be expected and are also not visible in the EDC's, indicating that states of various symmetries may be present and the simple model is therefore only a rough approximation. Nevertheless, besides the discussed strange polarisation effects near the M-point it describes very well the results of our Fermi surface maps of Bi-2201 and Bi-2212 taken with synchrotron radiation. It should be finally mentioned that all discrepancies between Fermi surface maps taken with synchrotron radiation and non-polarized HeI-light could be explained by a detailed analysis of the polarization effects. Therefore, the present investigation re-confirms the presence of a hole-like Fermi surface for the Bi-based cuprates. The peculiar polarisation effects together with the variety of extra structures like DR and SFS lead to a complex interplay and may be the reason for a misleading interpretation of the photoemission spectra towards an electron-like Fermi surface for the cuprates [4].

We would like to thank D. Kaiser for enormous help in crystal growth and Dr. S. Rogaschewski, Dr. P. Schäfer and B. Stoye for characterization of the samples. We gratefully acknowledge assistance by the staff of HASYLAB, namely Dr. P. Gürtler, and BESSY. The work received support by the German Ministry of Science and Technology under contract no. BMBF 05 SB8KH1 0 and BMBF 05 SB8 FKB.

**FIGURE CAPTIONS**

**Fig. 1** (a) Fermi surface map of $Bi_2Sr_{2-x}La_xCuO_{6+\delta}$ with x=0.40 at hν=18eV. A theoretical Fermi surface expected for n=2 Bi-2212 proposed by Ding et al. [2] is also shown. The Fermi surface due to the main band (MB) is drawn with thick lines. Additional surfaces are due to diffraction replica (DR) of the main band (dotted lines), diffracted by m**q**=(0.21π,0.21π) with the diffraction order *m*, and shadow bands (SFS) of the main band (dashed line) shifted by a vector (π,-π). **E** assigns the electrical field vector of the synchrotron radiation. (b) Typical spectra series of optimally doped $Bi_2Sr_{2-x}La_xCuO_{6+\delta}$. Shown is the ΓY-direction of the Brillouin zone, i.e. in the left panel a traditional energy distribution curve (EDC) series for hν=20eV and in the right panel a momentum distribution curve (MDC) series for hν=18eV. The polarization geometry is the same as in Figure 1a. The EDC plot reveals besides the Fermi level crossing of the main band at 0.4ΓY also crossing points of the diffraction replica near 0.21ΓY and 0.62ΓY. In the MDC plot, in addition, the shadow Fermi surface (SFS) can be clearly resolved from the diffraction replica (DR) surface around 0.65ΓY. It should be noted that different samples and photon energies were used for the EDC and MDC series.

**Fig. 2** Left: Experimental Fermi surface map of $Bi_2Sr_2CaCu_2O_{8+\delta}$ at hν=18eV. **E** assigns the electrical field vector of the synchrotron radiation. The theoretical Fermi surface expected for Bi-based cuprates [2] is the same as in Figure 1a. Right: Left: Experimental Fermi surface map of $Bi_2Sr_2CaCu_2O_{8+\delta}$ at hν=18eV. **E** assigns the electrical field vector of the synchrotron radiation. The theoretical Fermi surface expected for Bi-based cuprates [2] is the same as in Figure 1a. Right: Fermi surface map calculated for the same polarization conditions for the region marked by the white dotted line in the left panel, using a concept given by Mesot et al. [27].

**Fig. 3** Experimental geometries applied experimentally for the discussion of the polarisation effects. In the upper panel the ΓX(Y) planes are parallel to the electrical field vector **E** of the linearly polarized synchrotron light, in the lower panel the ΓM plane. For the Bi-based cuprate superconductors an initial state of d $x^2{-}y^2$ orbital symmetry is assumed.

**Fig. 4** Normal state spectra (T=40K) of optimally doped $Bi_2Sr_{2-x}La_xCuO_{6+\delta}$ at hν=18 eV. The spectra have been taken on the Fermi surface, i.e. along the directions ΓX (upper panel) and ΓY (lower panel) of the Brillouin zone for two polarisation geometries. For $X_1$ and $Y_1$, the electrical field vector is parallel to the detection plane (**E** ∥ **p**), for $X_2$ and $Y_2$ **E** is perpendicular to **p** (see insets). Note that between the two spectra of each panel the sample has been turned by 90°. The intensities are absolutely normalized by the incoming photon flux.

**Fig. 5** Normal state spectra (T=35K) of optimally doped $Bi_2Sr_{2-x}La_xCuO_{6+\delta}$ at the M-point for different polarisation geometries (hν=34 eV). Upper panel (clean polarisation geometry): The sample is kept fixed and the electron detector has been moved to the M-points M(π,0) and M(0,π). Lower panel (mixed polarisation geometry): Here the sample has again been kept fixed and the direction of the electrical field vector **E** of the synchrotron light has been turned by 90 °. Note that for these two geometries the transition matrix element $M_{f,i}$ is unchanged. The intensities are absolutely normalized by the incoming photon flux.

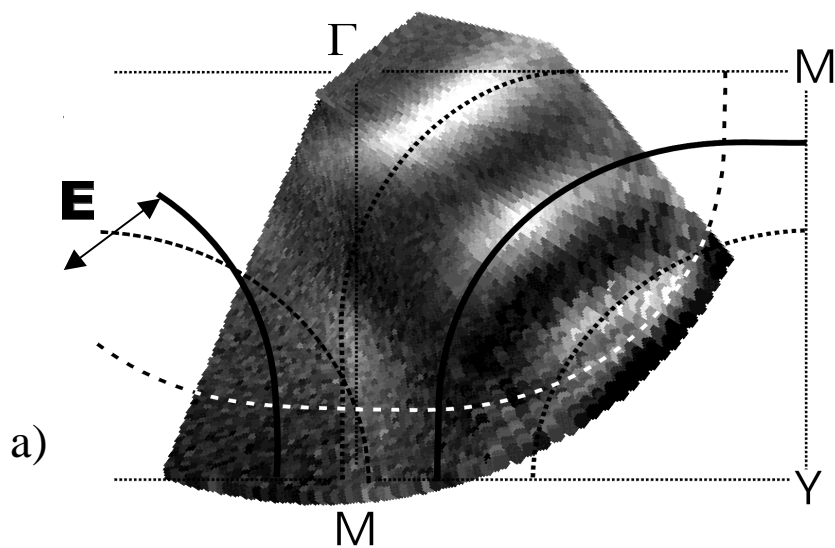

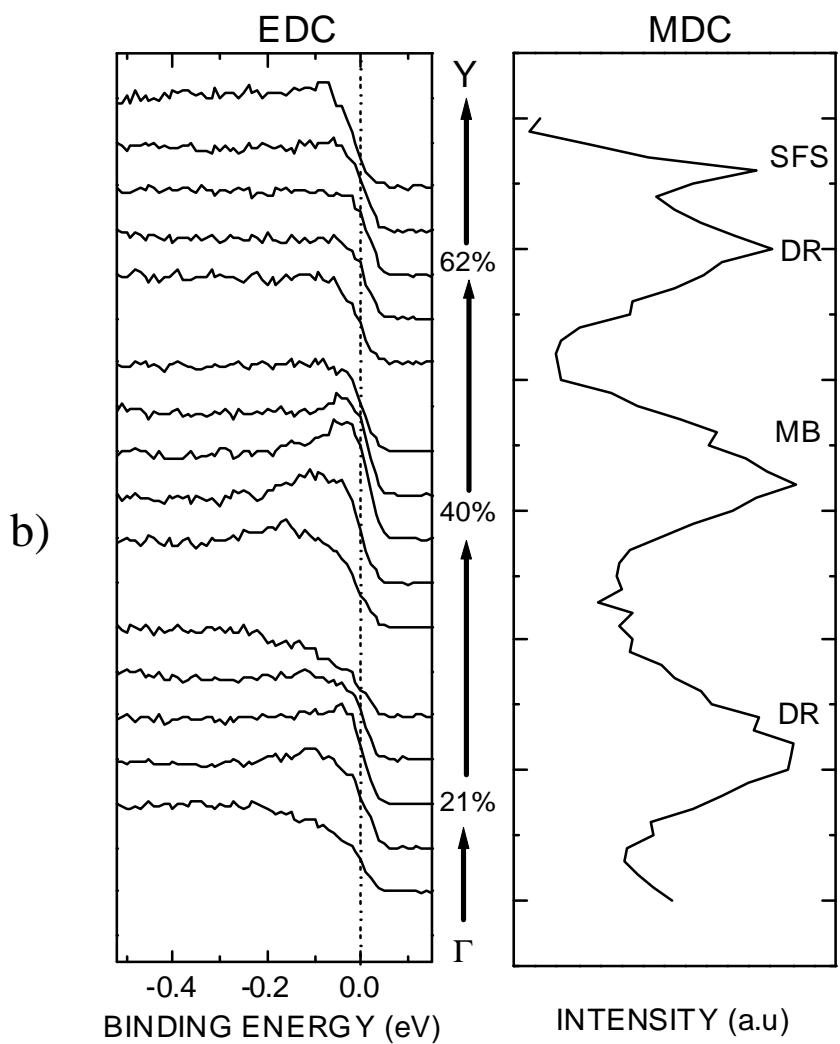

Figure 1
R Müller et al.

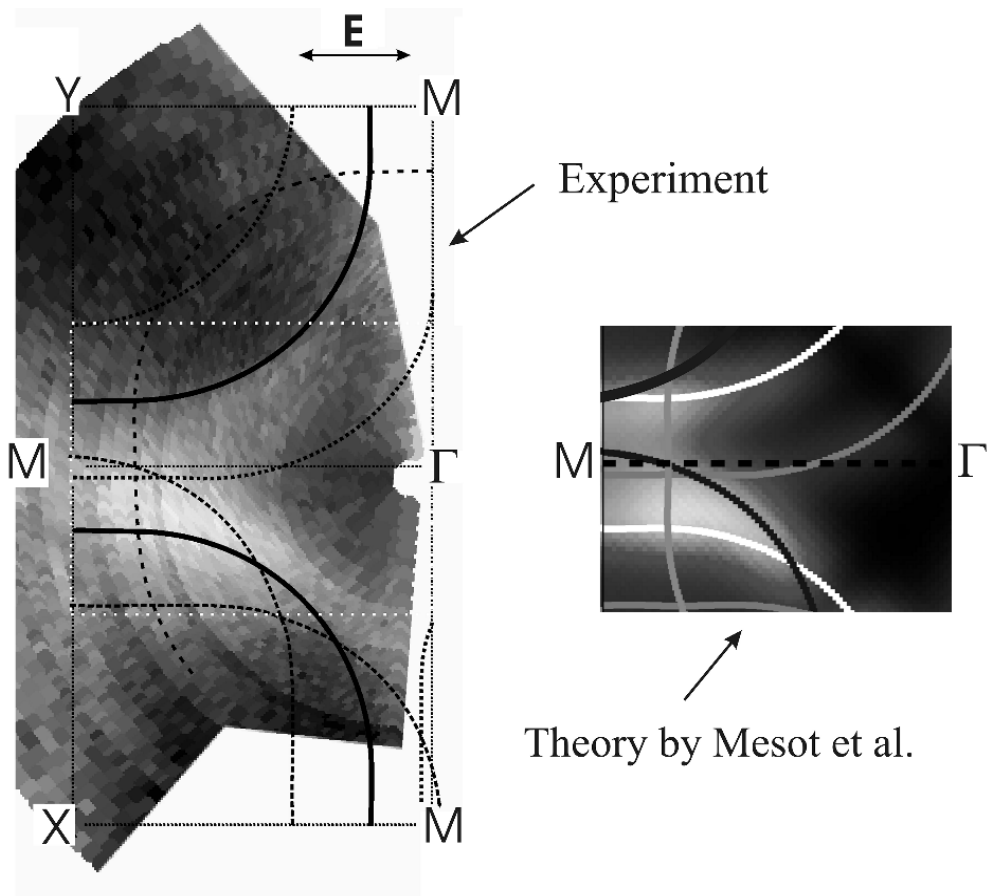

Figure 2
R.Müller et al.

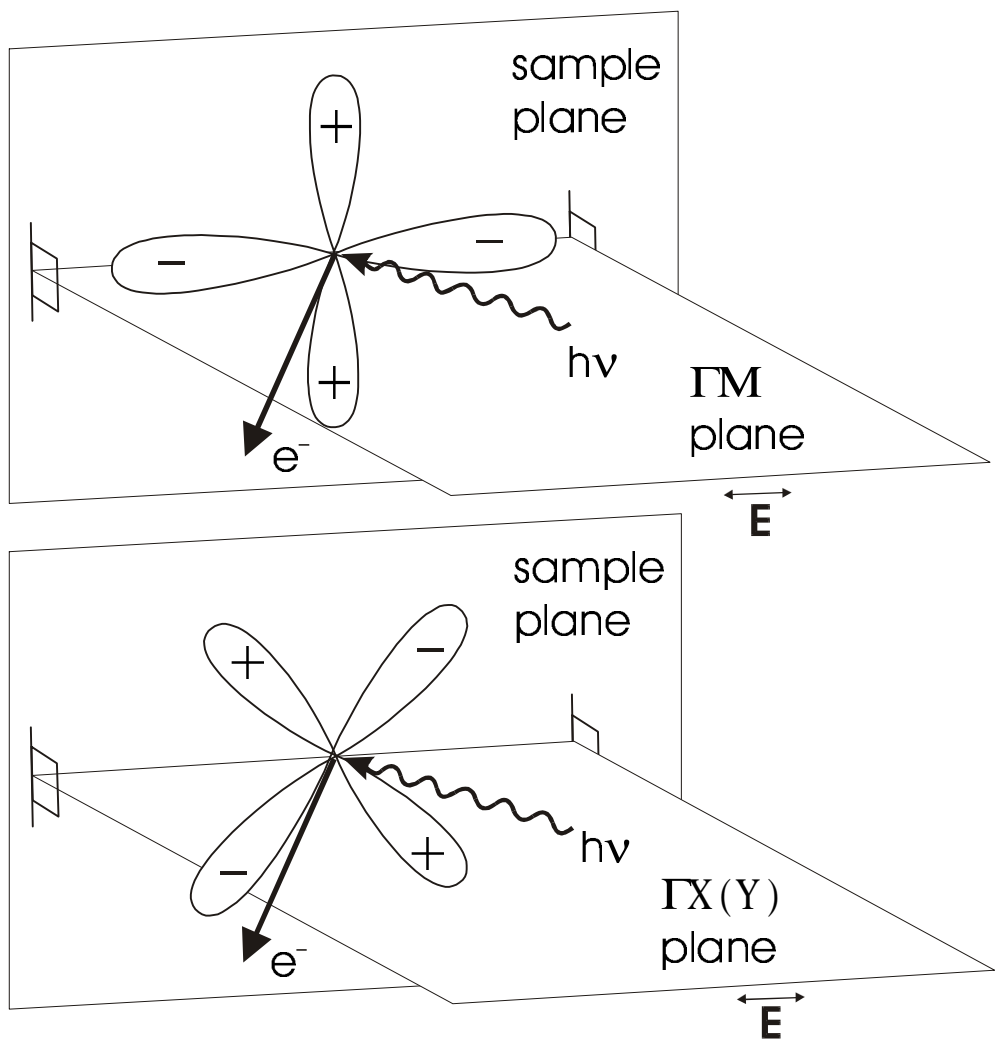

**Figure 3**
R. Müller et al.

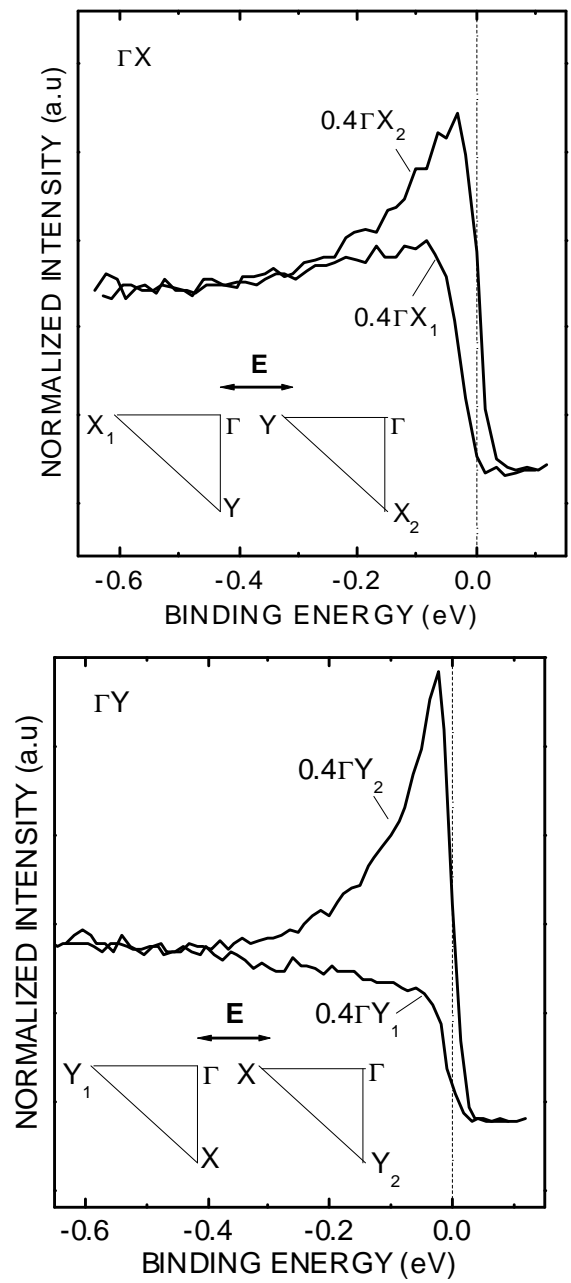

**Figure 4**
R. Müller et al.

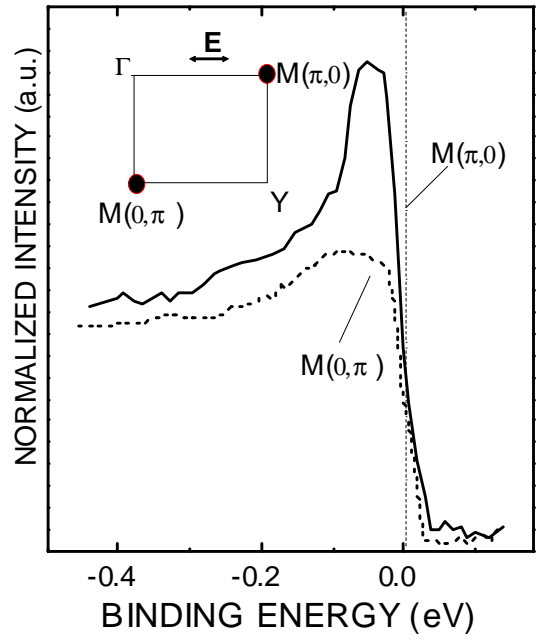

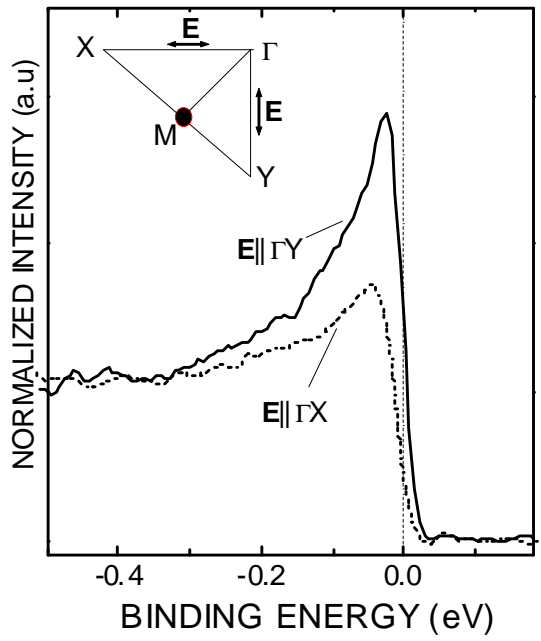

**Figure 5**
R. Müller et al.